# Nodeless superconductivity in the cage-type superconductor Sc$_5$Ru$_6$Sn$_{18}$ with preserved time-reversal symmetry

**D. Kumar†[1], C. N. Kuo[1], F. Astuti[2,3], T. Shang[4,5,6], M. K. Lee[1], C. S. Lue[1], I. Watanabe[2,3], J. A. T. Barker[7], T. Shiroka[8,9], L. J. Chang*[1,10]**

[1]Department of Physics, National Cheng Kung University, Tainan - 70101, Taiwan.

[2]RIKEN Nishina Center, Wako, Saitama 351-0198, Japan.

[3]Department of Condensed Matter Physics. Graduate School of Science, Hokkaido University, Sapporo 060-0810, Japan.

[4]Laboratory for Multiscale Materials Experiments, Paul Scherrer Institut, Villigen CH-5232, Switzerland

[5]Swiss Light Source, Paul Scherrer Institut, Villigen CH-5232, Switzerland

[6]Institute of Condensed Matter Physics, École Polytechnique Fédérale de Lausanne (EPFL), Lausanne CH-1015, Switzerland

[7]Laboratory for Muon-Spin Spectroscopy, Paul Scherrer Institut, CH-5232 Villigen PSI, Switzerland

[8]Paul Scherrer Institut, CH-5232 Villigen PSI, Switzerland

[9]Laboratorium für Festkörperphysik, ETH Zürich, CH-8093 Zurich, Switzerland

[10]Advanced Science Research Center, Japan Atomic Energy Agency (JAEA), Tokai, Naka, Ibaraki 319-1195, Japan.

†dinidixit@gmail.com, *ljchang@mail.ncku.edu.tw

## Abstract

We report the single-crystal synthesis and detailed investigations of the cage-type superconductor Sc$_5$Ru$_6$Sn$_{18}$, using powder x-ray diffraction (XRD), magnetization, specific-heat and muon-spin relaxation (μSR) measurements. Sc$_5$Ru$_6$Sn$_{18}$ crystallizes in a tetragonal structure (space group *I4$_1$/acd*) with the lattice parameters *a* = 1.387(3) nm, *c* = 2.641(5) nm. Both DC and AC magnetization measurements prove the type-II superconductivity in Sc$_5$Ru$_6$Sn$_{18}$ with $T_c$ ≈ 3.5(1) K, a lower critical field $H_{c1}(0)$ = 157(9) Oe and an upper critical field, $H_{c2}(0)$ = 26(1) kOe. The zero-field electronic specific-heat data are well fitted using a single-gap BCS model, with Δ(0) = 0.64(1) meV. The Sommerfeld constant γ varies linearly with the applied magnetic field, indicating *s*-wave superconductivity in Sc$_5$Ru$_6$Sn$_{18}$. Specific-heat and transverse-field (TF) μSR measurements reveal that Sc$_5$Ru$_6$Sn$_{18}$ is a superconductor with strong electron-phonon coupling, with TF-μSR also suggesting the single-gap *s*-wave character of the superconductivity. Furthermore, zero-field μSR measurements do not detect spontaneous magnetic fields below $T_c$, hence implying that time-reversal symmetry is preserved in Sc$_5$Ru$_6$Sn$_{18}$.

## 1. Introduction

In 1957, Bardeen, Cooper, and Schrieffer (BCS) explained superconductivity by using the concept of Cooper pairs [1], implying electrons with equal and opposite spins and crystal momenta, which pair together. Understanding the pairing mechanism in unconventional superconductors is a challenging task. In conventional *s*-wave superconductors, only the gauge symmetry is broken. However, in case of unconventional pairing, besides the global gauge symmetry (responsible for the Meissner- and Josephson effects [2]) other symmetries of the Hamiltonian might be broken in the superconducting state, including spin-rotation, lattice-point and translation group symmetries. Studying such broken symmetries in superconductors is crucial and it can be achieved by investigating the symmetry properties of the order parameter, $\psi(k)$. Depending on the parity of superconducting order parameter [3], superconductors with an inversion center may be classified in either even parity spin-singlet ($S = 0$) or in odd parity spin-triplet pairing states ($S = 1$). For instance, a few compounds, such as the 4*d*-electron system $Sr_2RuO_4$ [4,5] or the 5*f*-electron system $UGe_2$ [6] have been reported to be spin-triplet superconductors.

Besides modifying the properties of the system, broken symmetry may lead to some interesting unconventional behavior. Superconductivity itself is one of the best examples of a symmetry-breaking phenomenon. Time-reversal symmetry (TRS) breaking is another interesting example. TRS breaking is a rare phenomenon and has been observed only in a few unconventional superconductors, such as $Sr_2RuO_4$ [4], $LaNiC_2$ [7], $Re_6Zr$ [8], $Re_{24}Ti_5$ [9]. TRS breaking can be probed with the help of zero-field muon spin-relaxation (ZF-μSR) technique, by detecting the occurrence of tiny spontaneous magnetic fields, below the onset of superconductivity. The presence of such spontaneous fields restricts the pairing symmetry of the superconducting states. TRS breaking is associated with a special kind of superconducting states having a degenerate representation. The two or more degenerate superconducting phases lead to a spatially inhomogeneous order parameter near the resulting domain walls; which, in turn, create spontaneous supercurrents and hence, spontaneous magnetic field in those regions [3]. TRS breaking fields may also originate from the intrinsic magnetic moments due to spin-polarization (in case of spin-triplet pairing) and the relative spin-angular momentum associated with the Cooper pairs [10]. Recently, some of the noncentrosymmetric superconductors, such as $Re_6Zr$ [8] and $La_7Ir_3$ [11], with a mixed singlet-triplet pairing, were found to exhibit TRS breaking. However, it has been established theoretically [12] and experimentally [13] that the presence of singlet-triplet mixing not necessarily implies a broken TRS.

Compounds with cage-like structures have attracted remarkable attention due to their peculiar features. There are three major classes of cage-type compounds which are being studied extensively

*i.e.*, skutterudites ($RT_4X_{12}$), $\beta$-pyrochlore oxides ($AOs_2O_6$) and Ge/Si clathrates [14]. Exotic phenomena such as heavy-fermion superconductivity or exciton-mediated superconductivity were discovered in these materials. These compounds consist of three-dimensional skeletons which surround large atomic cages, in which small atoms are situated. Because of a strong electron-phonon coupling and to weak structural couplings, the small atoms can "rattle" with large atomic excursions, ultimately leading to a rattling vibration. Such rattling of small atoms might result in interesting phenomena, such as strong-coupling superconductivity in $AOs_2O_6$ [15]. A specific case of cage-type compounds is given by $R_5Rh_6Sn_{18}$ ($R$ = Sc, Y, Lu). These crystallize in a tetragonal structure with space group of $I4_1/acd$ and $Z$ = 8, where $Z$ represents the number of formula units per unit cell ($R$ occupies sites of different symmetry [16]). $R_5Rh_6Sn_{18}$ exhibit superconductivity at 5 K (Sc), 3 K (Y) and 4 K (Lu) [17], respectively. The superconducting properties of $Lu_5Rh_6Sn_{18}$ and $Y_5Rh_6Sn_{18}$ compounds have been studied [2,18]. Unconventional superconductivity has been observed in both $Lu_5Rh_6Sn_{18}$ and $Y_5Rh_6Sn_{18}$, where former has an isotropic superconducting gap, while later show anisotropic gap. In addition, ZF-μSR studies reveal the presence of spontaneous magnetic fields, hinting at TRS breaking. However, the superconductivity of ruthenate stannides $R_5Ru_6Sn_{18}$ is largely unexplored. This motivated us to study the superconducting properties of $Sc_5Ru_6Sn_{18}$ and to search for possible TRS breaking in this compound.

In this paper, we report on the superconducting properties of the cage-type superconductor $Sc_5Ru_6Sn_{18}$ investigated via magnetization, specific-heat, and μSR measurements. The symmetry of the superconducting gap was studied using TF-μSR, whereas ZF-μSR measurements could not detect the spontaneous magnetic fields below $T_c$, hence indicating that the TRS is preserved in $Sc_5Ru_6Sn_{18}$. We also report on the calculated critical current density ($J_c$) as obtained from the isothermal hysteresis loops in $Sc_5Ru_6Sn_{18}$.

**2. Experimental methods**

Single crystals of $Sc_5Ru_6Sn_{18}$ were grown using a Sn-flux method with Sc-powder (99.99%), Ru-powder (99.99%) and Sn-shot (99.99%) as the starting materials. The typical dimensions of the crystals used in our investigations were 2.5×2.8×2.5 mm$^3$ as shown in the inset of Fig. 1. The smaller crystals were crushed into powder for X-ray diffraction (XRD) measurements (using a Bruker AXS GmbH D2 Phaser desktop X-ray diffractometer) with Cu-K$_\alpha$ radiation. The quality of the single crystal was verified using a 2D XRD technique with omega scan without crystal rotation. Well-defined spots on the 2D image indicated a good crystalline quality. The magnetization was measured using a superconducting quantum interference device (SQUID) magnetometer (Quantum Design) at temperatures down to 1.8 K and magnetic fields up to 20 kOe. The specific-heat measurements were performed in various magnetic fields (up to 30 kOe) in the temperature range 1.8 – 10 K, using the heat-capacity option of a Physical Property Measurement System (PPMS) (Quantum Design). The temperature dependence of AC susceptibility was again studied with PPMS, using a small (5 Oe) ac-driving field with frequencies up to 10 kHz.

The ZF-µSR measurements were carried out at the pulsed muon beam of the RIKEN-RAL Muon Facility at ISIS (United Kingdom). In this case the sample temperature was varied from about 30 K ($> T_c$) down to 1.5 K ($< T_c$) with cooling being performed in a helium-flow type cryostat (Janis Co.). Helium exchange gas was used to achieve good temperature homogeneity. The asymmetry parameter is defined as $A(t) = [F(t) - \alpha B(t)]/[(F(t) + \alpha B(t)]$, where $F(t)$ and $B(t)$ represent the muon events recorded in the forward and backward counters, respectively. $\alpha$ is a geometrical factor, which accounts for the different solid angles and efficiencies of the two detectors, as viewed from the sample position. The time dependence of $A(t)$, also known as the µSR time spectrum, was measured. The TF-µSR measurements were carried out in the superconducting mixed state in an applied field of 300 Oe, using the General-Purpose Surface (GPS) muon instrument located at the πM3 beamline of the Swiss Muon Source of the Paul Scherrer Institute in Villigen, Switzerland.

## 3. Results and discussion

### 3.1 Crystal structure

Figure 1 shows the room temperature powder XRD pattern for $Sc_5Ru_6Sn_{18}$ with the scattering angle $2\theta$ varying between 20º – 80º. All the peaks in the pattern can be well indexed using a tetragonal structure with lattice constants $a = 1.387(3)$ nm, $c = 2.641(5)$ nm, which give a unit cell volume = $5.080(5)$ nm$^3$ and a density $\rho = 7.8(3)$ g/cm$^3$ (space group $I4_1/acd$.) The obtained values are in good agreement with those previously reported on the $R_5M_6Sn_{18}$ ($R$ = rare earth, $M$ = transition metal) family of compounds [19, 20].

### 3.2 Magnetization

Figure 2 shows the hysteresis curves of $Sc_5Ru_6Sn_{18}$ recorded at various temperatures below $T_c$, characterized by symmetric *M-H* loops. The *M (H)* curves exhibit a butterfly shape, typical of type-II superconductors. The symmetry of the hysteresis loop (or the lack of it) allows one to distinguish between pinning- and surface- (or geometrical barrier) induced hysteresis, with flux pinning known to produce symmetric hysteresis loops [21], as in our case. The lower- $H_{c1}$ and upper- $H_{c2}$ field values were determined from the magnetization data. The lower critical field, $H_{c1}$, is defined as the point where the field-dependent magnetization starts to deviates from linearity (see inset in Fig. 3) [22]. Since the actual magnetic field around the sample is larger than the applied magnetic field due to the demagnetizing effects by a factor of $1/(1-N)$, where $N$ is the demagnetizing factor, so we have to correct the $H_{c1}$ values for that effect [23]. We have estimated the value of $N$ from the rectangular prism approximation which is based on the dimensions of the crystal [24] and we get $N = 0.39$. The main panel of Fig. 3 shows the temperature variation of corrected values of $H_{c1}$, which can be fitted by the relation [25]:

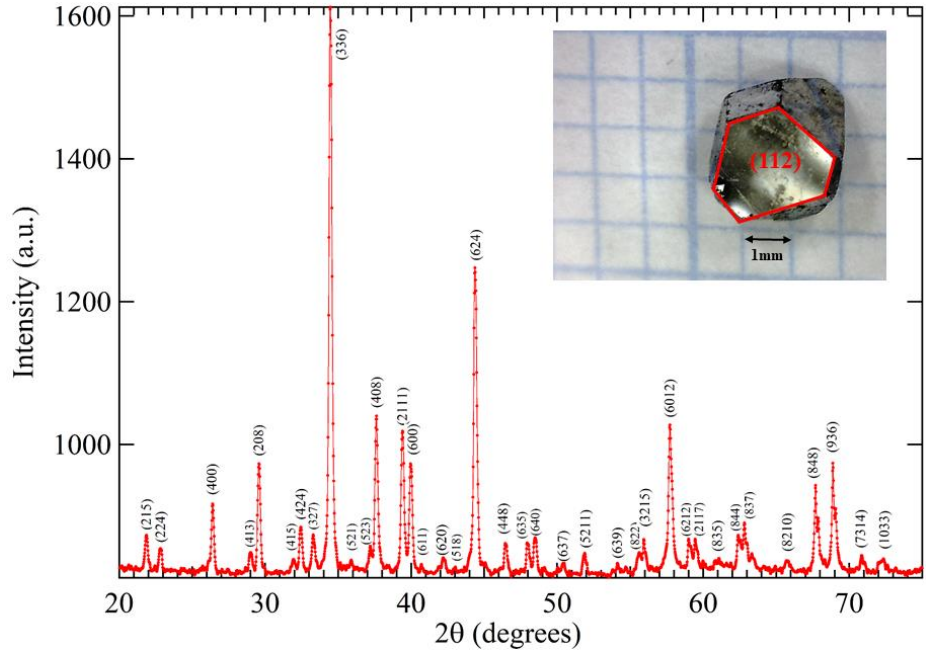

FIG. 1: Powder XRD pattern of $Sc_5Ru_6Sn_{18}$. All the peaks were indexed successfully to a tetragonal $R_5M_6Sn_{18}$ ($R$ = rare earth, $M$ = transition metal) phase. The inset shows the single crystal used in our study. The hexagonal-shaped plane was identified with (112) plane by means of X-ray reflection.

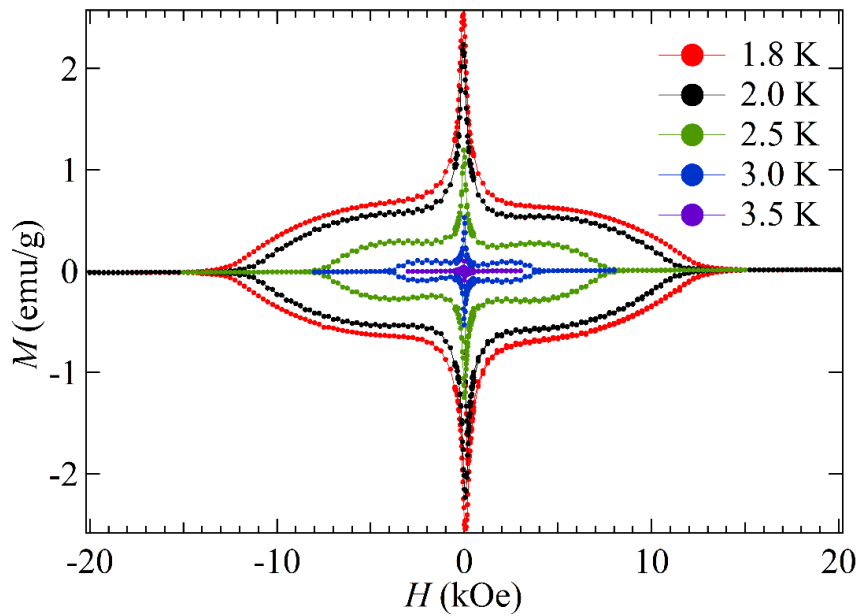

FIG. 2: Magnetization hysteresis curves of a $Sc_5Ru_6Sn_{18}$ crystal, recorded at different temperatures in an applied magnetic field normal to the (112) plane.

$$H_{c1}(T) = H_{c1}(0)\left[1-\left(\frac{T}{T_c}\right)^2\right] \quad (1)$$

where $H_{c1}(0)$ is the lower critical field at zero-temperature. The fit gives $H_{c1}(0) = 157(9)$ Oe.

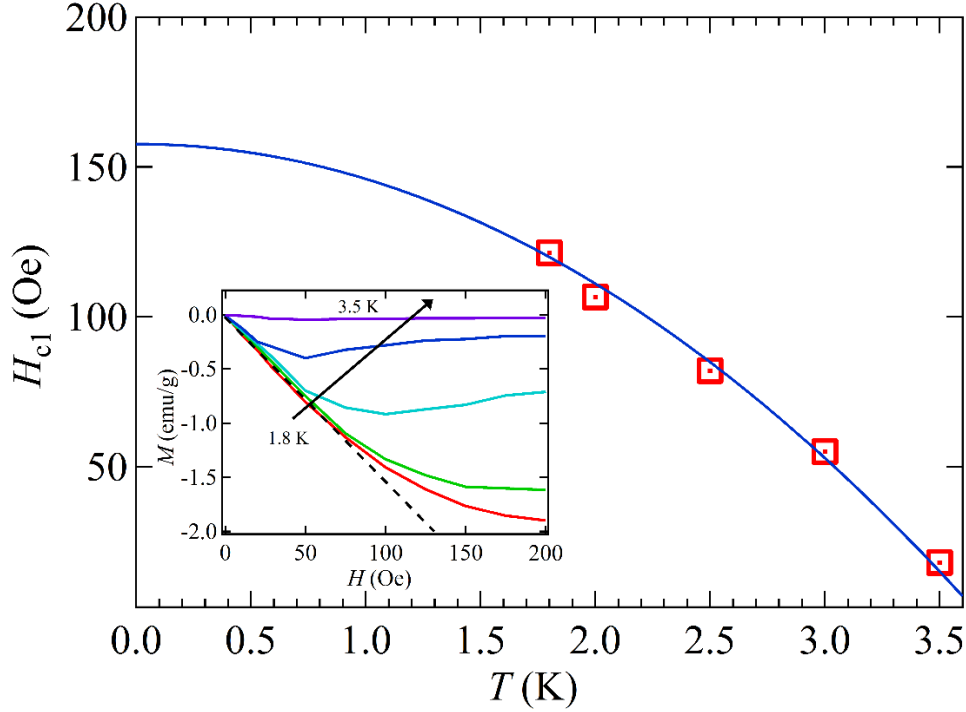

FIG. 3: Temperature dependence of $H_{c1}$ for $Sc_5Ru_6Sn_{18}$ (squares) and fit to equation (1) (solid line). Inset: Magnetization curves for $Sc_5Ru_6Sn_{18}$ for temperatures in the 1.8 – 3.5 K range. $H_{c1}$ is defined as the field where $M(H)$ curve starts to deviate from the linearity.

The upper critical field is defined as the point where the magnetization curve touches the zero magnetic moment line (see inset in Fig. 4). The main panel in Fig. 4 shows the temperature variation of $H_{c2}$, which can be fitted using the Ginzburg-Landau equation given by [25]:

$$H_{c2}(T) = H_{c2}(0)\frac{(1-q^2)}{(1+q^2)} \quad (2)$$

where $q = \frac{T}{T_c}$ and $H_{c2}(0)$ is the upper critical field at zero-temperature. The fit gives $H_{c2}(0) = 26(1)$ kOe. From this, we estimate a Ginzburg-Landau coherence length, $\xi(0) \approx 11.26(3)$ nm, by using the relation [25]:

$$H_{c2}(0) = \frac{\phi_0}{2\pi\xi^2(0)} \quad (3)$$

where $\varphi_0 = 2.07\times10^{-15}$ Tm$^2$, is the quantum of the magnetic flux. In addition, we can estimate the Ginzburg-Landau penetration depth using the relation [25]:

$$H_{c1}(0)= \frac{\phi_0}{2\pi\lambda^2(0)}\left[\ln\frac{\lambda(0)}{\xi(0)} + 0.12\right] \quad (4)$$

By using $H_{c1}(0)$ = 157(9) Oe and $\xi(0) \approx 11.26(3)$ nm, we obtain $\lambda(0)$ = 260(7) nm.

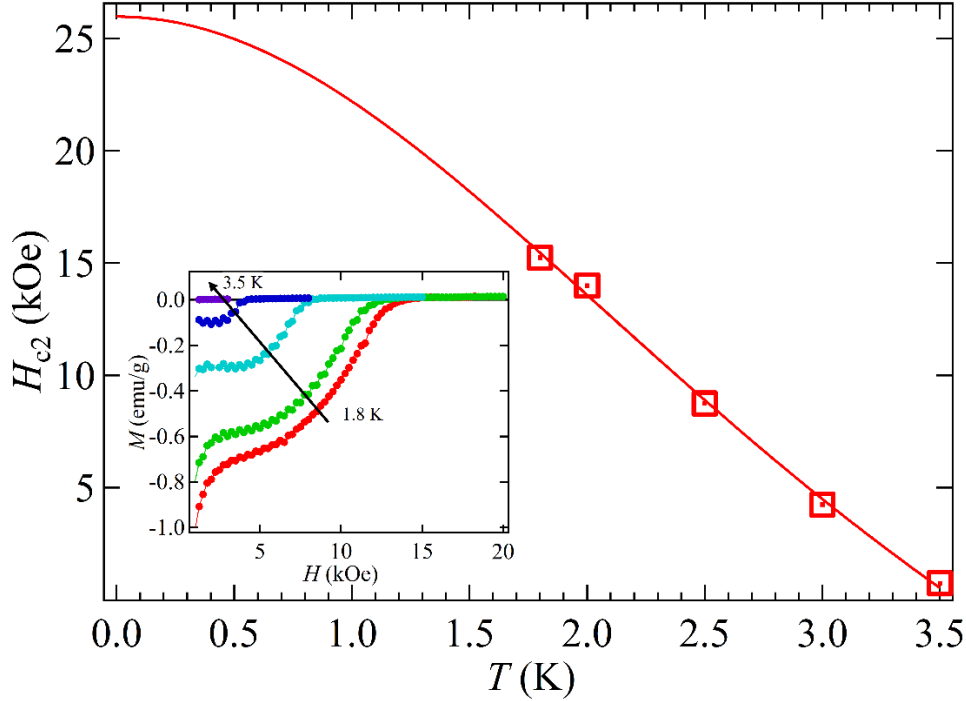

FIG. 4: Temperature dependence of $H_{c2}$ for $Sc_5Ru_6Sn_{18}$ (squares) and a fit to equation (2) (solid line). Inset: Magnetization curves for $Sc_5Ru_6Sn_{18}$ for temperatures in the 1.8 – 3.5 K range. $H_{c2}$ is defined as the field at which the magnetization becomes zero.

Finally the Ginzburg-Landau parameter ($\kappa$) is obtained from the relation $\kappa = \frac{\lambda(0)}{\xi(0)}$. By using the values for $\lambda(0)$ and $\xi(0)$, we find, $\kappa \approx 23(2) > \frac{1}{\sqrt{2}}$, the latter indicating that $Sc_5Ru_6Sn_{18}$ is indeed a strong type-II superconductor. The zero-temperature thermodynamic critical field was estimated by using relation [25],

$$H_{c2}(0) = \sqrt{2}\kappa H_c(0) \quad (5)$$

By using the values for $H_{c2}(0)$ = 26(1) kOe and $\kappa \approx 23(2)$, we get $H_c(0)$ = 799(6) Oe.

The field dependence of critical current density ($J_c$) was derived from the width $\Delta M$ of the magnetization curve, by using the Bean model [26, 27]:

$$J_c = 20\Delta M/[d(1-d/3b)] \quad (6)$$

where $d$ (mm) and $b$ (mm) are the sample dimensions ($d < b$). In our case $d$ = 1.25 mm and $b$ = 1.40 mm. The critical current density ($J_c$) is expected to have a maximum at the lower critical field, whereas above this threshold it decreases rapidly with the increasing field. Fig. 5(a) shows $J_c$ as a function of the magnetic field at various temperatures. Above the threshold field, for each isothermal measurement, initially $J_c$ shows an exponential decrease followed by a power-law variation, in good agreement with the previous reports [28,29]. We obtain $J_c(1.8\ \text{K}) \sim 6(3) \times 10^8$ A/m². The de-pairing current is given by the relation:

$$J_d = \frac{\varphi_0}{3\sqrt{3}\mu_0\pi\lambda^2\xi} \tag{7}$$

By using the previously obtained values of $\xi(0) = 11.26(3)$ nm, $\lambda(0) = 260(7)$ nm, we find $J_d = 3.07(5) \times 10^{10}$ A/m². The long coherence length implies a high pinning energy and flux-lines that do not move easily [30]. The material might have a large concentration of weak pinning centers, which ultimately leads to collective pinning. In the scenario of collective pinning, the critical current density depends strongly on the magnetic field and is typically small. This behavior has been observed *e.g.*, in case of layered inhomogeneous superconductors [31]. The irreversibility field ($H_{irr}$), at which the magnetic hysteresis disappears, is determined by the criterion $J_c = 2 \times 10^7$ A/m² [28]. The variation of the $H_{irr}$ and $H_{c2}$ with temperature is shown in Fig. 5(b). We note that the irreversibility field is comparable to the upper critical field ($H_{c2}$), a rather plausible result considering that the material has a low $T_c$ and a high coherence length [30].

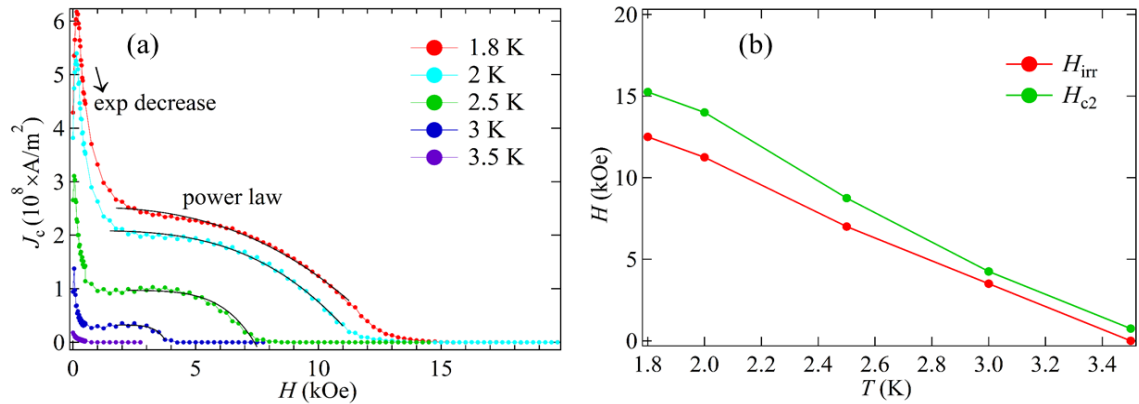

FIG. 5: (a) Variation of the critical current density ($J_c$) with the applied magnetic field for different temperatures for single crystal of $Sc_5Ru_6Sn_{18}$. The critical current density shows a peak near the lower critical field ($H_{c1}$) and ensuing exponential decrease followed by a power law ($H^a$) variation vs.applied field. The $J_c(1.8\ \text{K}) \sim 6 \times 10^8$ A/m² is obtained at a low field of 150 Oe. (b) Phase diagram showing the variation of the irreversibility ($H_{irr}$) and upper critical ($H_{c2}$) fields for $Sc_5Ru_6Sn_{18}$.

### 3.4 AC susceptibility

To further confirm the superconducting transition temperature we measured the ac magnetic susceptibility in the temperature range 1.8 – 10 K, with $H_{ac} = 5$ Oe and $H_{dc} = 0$ Oe. Fig. 6(a) shows the imaginary part of the ac susceptibility, while Fig. 6(b) shows the real part, corresponding to the out-of-phase and in-phase component, respectively (with respect to the ac field). The real part of the ac susceptibility vs temperature, represents the transition from the Meissner state (perfect shielding) to the complete penetration of the ac magnetic field inside the sample. On the other hand, the imaginary part of the ac susceptibility represents the ac losses occurring when the ac field penetrates the sample. As shown in Fig. 6(a), $\chi''$ has a sharp transition near $T_c$ = 3.6 K. The transition in $\chi''$ is independent of the frequency of the driving ac field. The appearance of a peak in $\chi''$ is commonly associated with the superconducting transition. When the temperature is far below the critical temperature and the magnetic field is smaller than the lower critical field $H_{c1}$, the screening current generated by the ac field is confined to regions near the sample surface. Hence, no magnetic flux enters the sample and the ac losses are minimal; consequently $\chi''$ is almost zero. As the temperature is raised, the magnetic field starts penetrating the sample and ac losses start increasing. The maximum in $\chi''$ occurs when the ac field fully penetrates the sample.

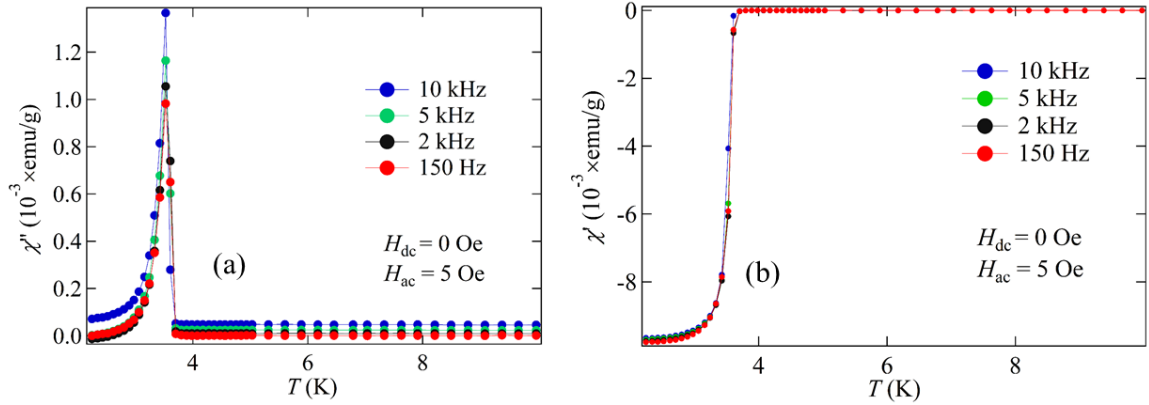

FIG. 6: (a) Temperature dependence of the imaginary part of ac susceptibility for $H_{dc}$ = 0 Oe and $H_{ac}$ = 5 Oe and different frequencies for $Sc_5Ru_6Sn_{18}$ (b) Temperature dependence of the real part of ac susceptibility measured in identical conditions.

### 3.3 Specific heat

The specific heat in the superconducting state is one of the key parameters to reflect closely the superconducting gap and its symmetry. Therefore, we measured and analyzed in detail the zero-field specific heat of $Sc_5Ru_6Sn_{18}$. The electronic specific heat ($C_e/T$) of the sample was obtained after subtracting the phonon contribution from the experimental data. As shown in the inset of Fig. 7, the normal-state specific heat was fitted using the relation $C/T = \gamma + \beta T^2 + \delta T^4$ for 14 K$^2 \leq T^2 \leq$ 42 K$^2$ to obtain $\gamma$ = 36.93(6) mJ/mol-K$^2$, $\beta$ = 2.5(5) mJ/mol-K$^4$ and $\delta$ = 1.2(9) mJ/mol-K$^6$. The Debye temperature $\theta_D$ = 205(1) K was calculated using the relation:

$$\beta = p(\frac{12}{5})\pi^4 R \theta_D^{-3} \tag{8}$$

where $p = 29$ is the number of atoms in one formula unit and $R = 8.314$ J/K-mol. The density of states at the Fermi level $N(E_F)$ was estimated from [32]:

$$\gamma = \frac{(\pi k_B)^2}{3} N(E_F) \tag{9}$$

where $k_B$ is the Boltzmann constant. By using the previously obtained $\gamma$ value, we obtain $N(E_F) = 15.24(6)$ states/eV per formula unit. The strength of electron-phonon coupling can be estimated by the McMillan equation [33]:

$$\lambda_{el-ph} = \frac{1.04 + \mu^* \ln\left(\frac{\theta_D}{1.45 T_c}\right)}{(1 - 0.62\mu^*)\ln\left(\frac{\theta_D}{1.45 T_c}\right) - 1.04} \tag{10}$$

Here $\mu^* = 0.13$ is the Coulomb repulsion parameter. By inserting the values for $\theta_D = 205(1)$ K and $T_c = 3.5(1)$ K, we get $\lambda_{el-ph} = 0.64(4)$, which indicates that $Sc_5Ru_6Sn_{18}$ is a strongly coupled superconductor.

The calculated density of states $N(E_F)$ and the effective mass of quasiparticles $m^*$, depend on the many-body electron-phonon interactions. These quantities are related to the bare band-structure of density of states $N_{band}(E_F)$ and $m^*_{band}$ by the following relations [34]:

$$N(E_F) = N_{band}(E_F)(1 + \lambda_{el-ph}) \tag{11}$$

$$m^* = m^*_{band}(1 + \lambda_{el-ph}) \tag{12}$$

By using the values of $N(E_F)$ and $\lambda_{el-ph}$ in equation (11), we get $N_{band}(E_F) = 9.270(6)$ states/eV f.u. By using $m^*_{band} = m_e$ (the free electron mass), in equation (12), we obtain $1.644\, m_e$ for the mass of the quasiparticles.

The density of states can be used to estimate the Fermi velocity, $v_F$, which is related to $N(E_F)$ by [35],

$$v_F = (\pi^2 \hbar^3 / m^{*2} V_{f.u.}) N(E_F) \tag{13}$$

Where $\hbar$ = Planck's constant $/2\pi$, $V_{f.u.} = V_{cell}/2$ is the volume per formula unit. Using the values of $m^*$, $V_{cell}$ and $N(E_F)$, we get $v_F = 1.94(7) \times 10^7$ cm/s for $Sc_5Ru_6Sn_{18}$. We can further estimate the mean free path ($l$) of the superconducting carriers by using relation $l = v_F \tau$, where $\tau$ is the mean free scattering time and given by $\tau = m^*/n_s e^2 \rho_0$, where $\rho_0$ is the residual resistivity and $n_s$ is the superconducting carrier density given by $n_s = m^{*3} v_F^3 / 3\pi^2 h^3$, assuming a spherical Fermi surface [35]. Using the value of $v_F$, we get $n_s = 7.05(2) \times 10^{26}$ carriers/m$^3$. By combining these expressions we get [35],

$$l = 3\pi^2 \left(\frac{\hbar}{e^2 \rho_0}\right)\left(\frac{\hbar}{m^* v_F}\right)^2 \tag{14}$$

Putting $\rho_0 = 200$ μΩ cm (resistivity data not shown here) and above estimated values of $m^*$ and $v_F$, we get $l = 8.14(5)$ nm.

The normalized electronic specific heat $C_e/\gamma T$ is plotted vs the normalized temperature $T/T_c$ as shown in Fig. 7. The normalized specific heat jump at $T_c$ is $\frac{\Delta C_e}{\gamma T_c} = 1.6$ for $\gamma = 36.93(6)$ mJ/mol-K$^2$. Since such value is higher than the BCS value of 1.43 (for a weakly coupled superconductor), it again indicates a strong electron-phonon coupling in Sc$_5$Ru$_6$Sn$_{18}$.

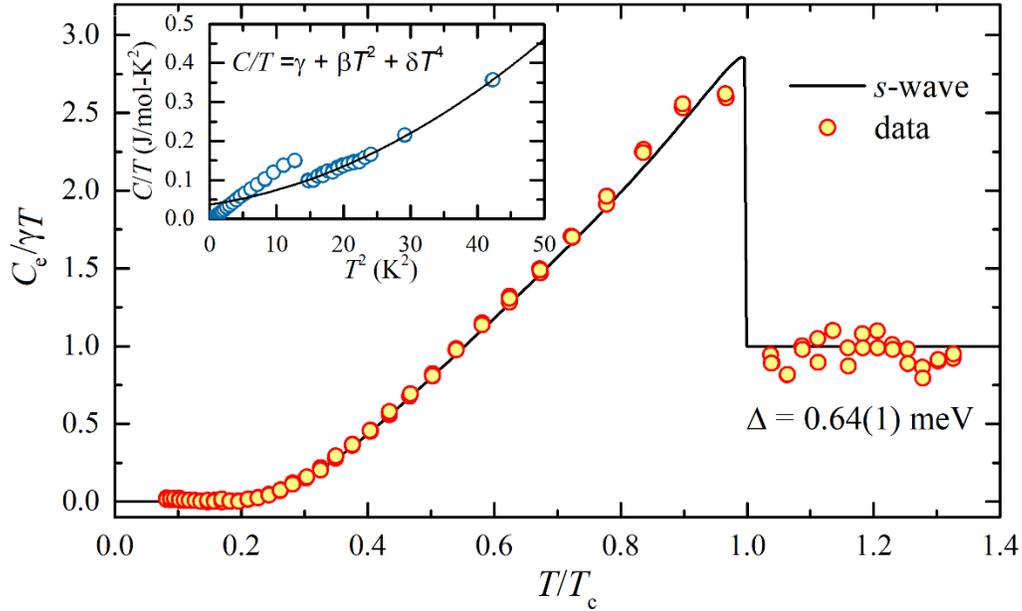

FIG. 7: Temperature dependence of the zero-field electronic specific heat of Sc$_5$Ru$_6$Sn$_{18}$. The solid line refers to a fit using a fully-gapped *s*-wave model. Inset: Raw $C/T$ data versus $T^2$. The solid line shows the fit to $\frac{C}{T} = \gamma + \beta T^2 + \delta T^4$, which is used to calculate the phonon contribution to the specific heat.

The temperature dependence of specific heat in the superconducting state is best modelled by an *s*-wave single-gap BCS expression of the normalized entropy $S$ [34]:

$$\frac{S}{\gamma T_c} = -\frac{6}{\pi^2}\left(\frac{\Delta(0)}{k_B T_c}\right)\int_0^\infty [f \ln(f) + (1-f)\ln(1-f)]dx \tag{15}$$

where $f(\xi) = \left[\exp\left(\frac{E(\xi)}{k_B T}\right) + 1\right]^{-1}$ is the Fermi function, $E(\xi) = \sqrt{\xi^2 + \Delta^2(q)}$, where $\xi$ is the energy of electrons in the normal state, measured with respect to the Fermi energy, $x = \xi/\Delta(0)$,

$q = T/T_c$ and $\Delta(q) = \Delta(0) \tanh\left[1.82\left(1.018\left(\frac{1}{q}\right) - 1\right)^{0.51}\right]$ is the temperature dependence of the superconducting gap. The normalized electronic specific heat is given by the expression [34]:

$$\frac{C_e}{\gamma T_c} = q \frac{d}{dq}\left(\frac{S}{\gamma T_c}\right) \qquad (16)$$

The electronic specific heat ($C_e$) in the superconducting state is described by Eq. (16), while it is equal to $\gamma T_c$ in the normal state. The specific heat data were fitted to Eq. (16) as shown in the Fig. 7. The fit gives a superconducting gap, $\Delta(0) = 0.64(1)$ meV.

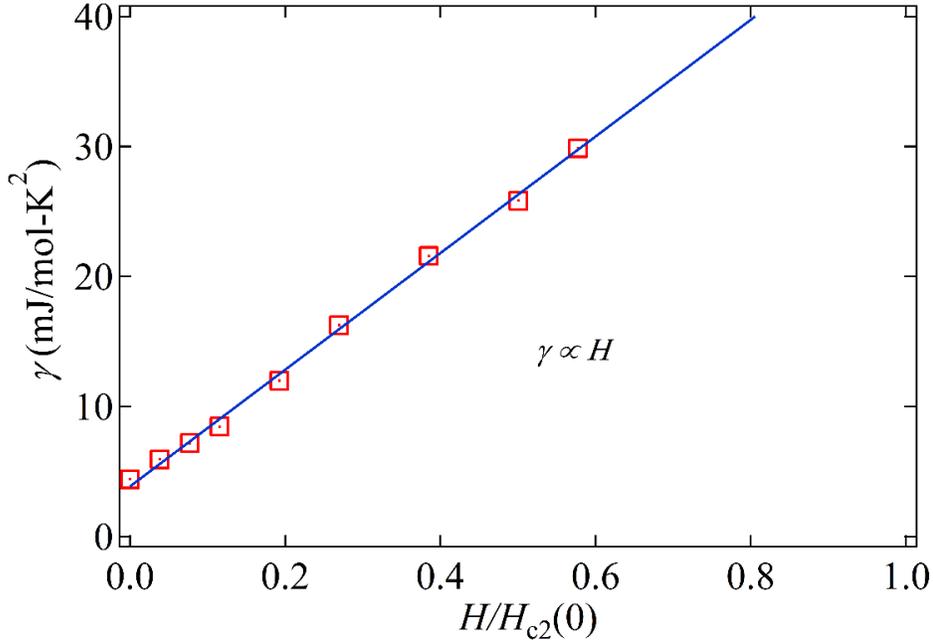

FIG. 8: Sommerfeld constant $\gamma$ vs normalized field $\frac{H}{H_{c2}(0)}$ for $Sc_5Ru_6Sn_{18}$. The solid line shows the linear relation between $\gamma$ and field, which indicates the $s$-wave nature of superconductivity in $Sc_5Ru_6Sn_{18}$.

The Sommerfeld constant ($\gamma$) is calculated by fitting the $\frac{C_e}{T}$ versus $T$ data for various fields using the relation [36]:

$$\frac{C_e}{T} = \gamma + \frac{A_1}{T}\exp\left(\frac{-A_2 T_c}{T}\right) \qquad (17)$$

where $A_1$ and $A_2$ are constants.

The single gap $s$-wave superconductivity in $Sc_5Ru_6Sn_{18}$ is also confirmed by the magnetic-field dependence of the electronic specific-heat coefficient (Sommerfeld constant), $\gamma(H)$. In a single-gap type-II superconductor, the Sommerfeld constant is proportional to the vortex density. When the applied magnetic field is increased, the density of vortices increases too, which, in turn, result in an increase of the quasiparticle density of states. Consequently, $\gamma$ turns out to be proportional to the applied magnetic field in case of a nodeless and isotropic s-wave superconductor [34,36,37]. On the other hand, in case of nodes in the superconducting gap, Volovik predicted a nonlinear relation given by $\gamma \propto H^{0.5}$ [38]. The field dependence of $\gamma$ is shown in Fig. 8. The linear relation between $\gamma$ and $H$ confirms the single-gap $s$-wave superconductivity in $Sc_5Ru_6Sn_{18}$.

The condensation energy $U(0)$ can be estimated from the relation:

$$U(0) = \frac{1}{2}\Delta^2(0)\, N_{band}(E_F) = \frac{3\gamma\Delta^2(0)}{4\pi^2 k_B^2} \tag{18}$$

Using the previously obtained values of $\gamma$ = 36.93(6) mJ/mol-K$^2$ and $\Delta(0)$ = 0.64(1) meV, we get $U(0)$ = 156.8(9) mJ/mol.

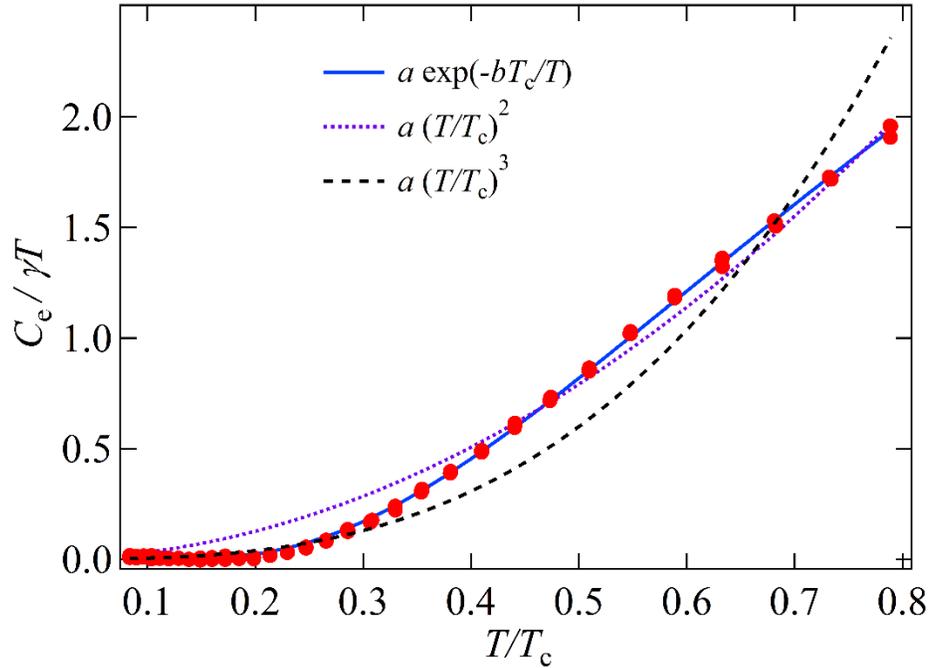

FIG. 9: Temperature dependence of the zero-field electronic specific heat for $Sc_5Ru_6Sn_{18}$ in the superconducting phase. The solid, dotted and dashed lines are fits to the data using three different models (see text for details).

To further confirm the characteristics of the superconducting gap function, we analyzed $C_e(T)$ by fitting the data by means of different functional forms *i.e.*, $e^{-b/T}$, $T^2$ and $T^3$, corresponding to the expected temperature dependence of a superconducting gap which is isotropic, has line nodes or linear point nodes, respectively. As can be seen in Fig. 9, well below $T/T_c$ = 0.8, data are best modelled by the exponential function, $\frac{C_e}{\gamma T} = ae^{-bT_c/T}$. The quadratic and cubic fits instead are rather poor. Similar exponential fits have been used to model the superconducting gap of other superconductors with comparable transition temperatures [9,39,40]. Thus, the above analysis confirms that $Sc_5Ru_6Sn_{18}$ has an isotropic gap and $s$-wave pairing.

We can estimate London penetration depth at $T$ = 0 K, $\lambda_L(0)$, using the relation [25]:

$$\lambda_L^2(0) = \frac{m^*c^2}{4\pi ne^2} = \frac{3\pi c^2 \hbar^3}{4m^{*2}e^2 v_F^3} \tag{19}$$

where $c$ is the speed of light in vacuum. Putting the values of all parameters, we get $\lambda_L(0) = 245(6)$ nm. The BCS coherence length $\xi_0$ can be estimated from $v_F$ and the superconducting energy gap $\Delta(0)$ using the relation [41]:

$$\xi_0 = \frac{\hbar v_F}{\pi \Delta(0)} \tag{20}$$

Putting the values of $v_F = 1.94(7) \times 10^7$ cm/s and $\Delta(0) = 0.64(1)$ meV, we get $\xi_0 = 63.72(9)$ nm. We find that BCS coherence length ($\xi_0$) is much larger than the mean free path $l = 8.14(5)$ nm, $l/\xi_0 = 0.12 \ll 1$, which indicates that the superconductivity in Sc$_5$Ru$_6$Sn$_{18}$ is in the moderate dirty limit.

### 3.5 µSR

Figure 10(a) shows the TF-µSR spectra collected above (5 K) and below (1.5 K) the superconducting transition temperature, $T_c$. The fast decay of muon-spin polarization below $T_c$ indicates an inhomogeneous field distribution due to the flux line lattice (FLL) in the vortex state. The time-domain spectra were fit by using the following model [9]:

$$G(t) = A_1 \cos(\gamma_\mu B_1 t + \Phi) e^{-\frac{(\sigma t)^2}{2}} + A_2 \cos(\gamma_\mu B_2 t + \Phi) \tag{21}$$

Here $A_1$ and $A_2$ are the initial muon-spin asymmetries, whereas $B_1$ and $B_2$ are the local fields sensed by muons implanted in the sample and the sample holder. $\gamma_\mu = 2\pi \times 135.53$ MHz/T is the muon gyromagnetic ratio, $\Phi$ is the phase factor and $\sigma$ is the Gaussian relaxation rate. Since the sample holder is non-magnetic, $B_2$ coincides with the applied magnetic field.

In the superconducting state, the Gaussian relaxation rate $\sigma$ contains contributions from both the FLL ($\sigma_{sc}$) and the nuclear magnetic moments ($\sigma_n$), such that:

$$\sigma^2 = \sigma_{sc}^2 + \sigma_n^2. \tag{22}$$

By using equation (22), one can estimate the FLL-related relaxation rate $\sigma_{sc} = \sqrt{\sigma^2 - \sigma_n^2}$, since $\sigma_n$ is expected to be temperature independent in the measured temperature range and is determined from the measurements made above $T_c$. Considering that $\sigma_{sc}$ is directly related to the superfluid density, the temperature dependence of $\sigma_{sc}$ provides hints on the superconducting gap and its symmetry. Further, $\sigma_{sc}$ can be modeled by [42,43,44]:

$$\frac{\sigma_{sc}(T)}{\sigma_{sc}(0)} = 1 + 2 \left\langle \int_{\Delta_k}^{\infty} \frac{E dE}{\sqrt{E^2 - \Delta_k^2}} \frac{\partial f}{\partial E} \right\rangle_{FS} \tag{23}$$

where *f* and *E* are the same as defined in equation (15). The curved brackets represent an average over the Fermi surface (FS). The superconducting gap is defined by $\Delta_k(q) = \Delta(q)g_k$, $(q = T/T_c)$, which contains an angular dependent part $g_k$ and a temperature dependent part $\Delta(q)$, which can be approximated as $\Delta(q) = \Delta(0) \tanh\left[1.82\left(1.018\left(\left(\frac{1}{q}\right) - 1\right)\right)^{0.51}\right]$. As shown in Fig. 10(b), the temperature dependence of the normalized FLL-related relaxation rate $\sigma_{sc}(T)/\sigma_{sc}(0)$, can be fitted by a single-gap isotropic *s*-wave using equation (23) (for single gap *s*-wave model, $g_k = 1$), with $\Delta(0) = 0.64(1)$ meV and $\sigma_{sc}(0) = 0.178 \mu s^{-1}$. This implies a ratio $2\Delta(0)/k_B T_c = 4.25(4)$ *i.e.*, higher than 3.53 expected for a weakly coupled BCS superconductor. This further confirms the strong electron-phonon coupling in $Sc_5Ru_6Sn_{18}$ and is consistent with the results obtained from the specific-heat data.

The effective penetration depth ($\lambda_{eff}$) for small applied fields, such that $H_{applied}/H_{c2} = 0.0115 \ll 1$, is given by relation [9,45]:

$$\frac{\sigma_{sc}^2(0)}{\gamma_\mu^2} = 0.00371 \frac{\varphi_0^2}{\lambda_{eff}^4} \tag{24}$$

Using $\sigma_{sc}(0) = 0.178 \mu s^{-1}$, we get $\lambda_{eff} = 774$ nm. The effective penetration depth ($\lambda_{eff}$) and London penetration depth $\lambda_L(0)$ are related by $\lambda_{eff} = \lambda_L(0)\sqrt{1 + \frac{\xi_0}{l}}$ [25]. Using $\lambda_{eff} = 774$ nm, $\xi_0 = 63.72(9)$ nm and $l = 8.14(5)$ nm, we get, $\lambda_L(0) = 258(7)$ nm, which is comparable to the value estimated using $m^*$ and $v_F$ and the magnetization measurements. This confirms the validity of our fitting model.

To probe the occurrence (or absence) of TRS breaking in $Sc_5Ru_6Sn_{18}$, we performed ZF- μSR experiments. The availability of 100% spin-polarized muon beams, along with the large muon gyromagnetic ratio, make ZF- μSR a very useful technique to detect the spontaneous internal fields, as has been shown in previous studies of $Y_5Rh_6Sn_{18}$ [18] and $Sr_2RuO_4$ [4]. In general, in absence of an external magnetic field, the onset of the superconducting phase does not induce any changes in the ZF-muon spin-relaxation rate. However, in case of TRS breaking, the presence of tiny spontaneous currents leads to the associated weak magnetic fields, which are detected by ZF- μSR as an increase in the muon spin-relaxation rate. Since the expected effects are rather small, we measured carefully the muon spin-relaxation rates both above and below $T_c$.

The ZF- μSR time-domain data for three representative temperatures (1.5, 15 and 30 K *i.e.*, below and above $T_c$) are shown in Fig. 11(a). Muons stopped in the silver sample holder give rise to a time independent background signal. Since no precession signals were observed in the entire 1.5 – 30 K temperature range, we rule out the possibility of significant internal fields and consequently, any long range magnetic order in $Sc_5Ru_6Sn_{18}$. In this scenario, the only possibility is that the muon-spin relaxation is due to the static, randomly distributed local fields, arising from the nuclear moments close to the muon-stopping site. In this case, the muon-spin depolarization can be described by the Kubo-Toyabe (KT) function:

$$G_{KT}(t) = \frac{1}{3} + \frac{2}{3}(1 - \sigma^2 t^2) \exp\left(\frac{-\sigma^2 t^2}{2}\right) \quad (25)$$

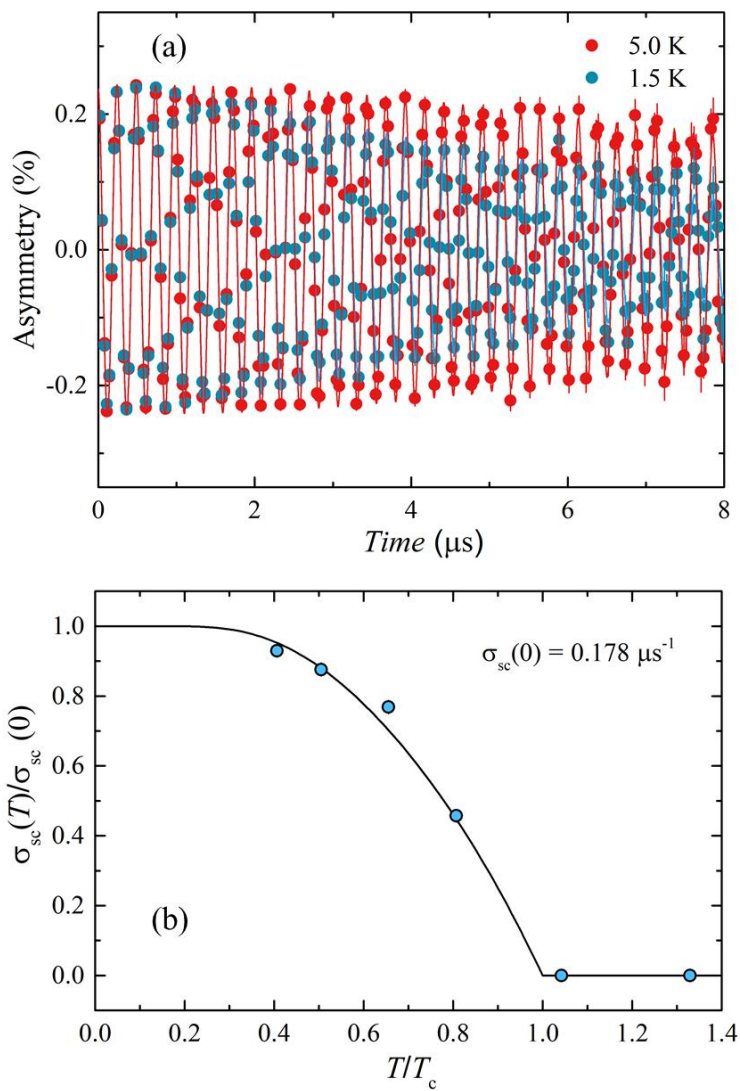

FIG. 10: (a) Transvers field µSR time spectra of $Sc_5Ru_6Sn_{18}$ measured at $T$ = 1.5 and 5 K in a magnetic field, $H$ = 300 Oe. The solid lines show fits to the data using equation (21). (b) Temperature dependence of the normalized muon spin-depolarization rate $\sigma_{sc}(T)/\sigma_{sc}(0)$ collected in the applied magnetic field, $H$ = 300 Oe. The solid line represents the fit by an isotropic $s$-wave superconducting gap using equation (23).

After subtracting the background due to the silver sample holder, the ZF-µSR spectra could be analyzed by using the function:

$$A(t) = A_0 G_{KT}(t) \exp(-\lambda_1 t) \qquad (26)$$

Here, $A_0$ is the initial asymmetry at $t = 0$. $G_{KT}$ can be treated as a temperature-independent parameter (within the µSR time window [46]). $\lambda_1$ is the muon-spin relaxation rate, considered to reflect the fluctuations of surrounding electronic spins lying close to the muon. Solid lines in Fig. 11(a) are the best-fit results by using equation (26). Fig. 11(b) shows the temperature dependence of $\lambda_1$, as obtained from the fits of the time spectra. Here $\lambda_1$ shows a slight enhancement with decreasing temperature, starting from $T > T_c$. This would mean that some unpaired electronic spins still remain, thus causing the observed muon-spin depolarization behavior.

A key result of the ZF-µSR data is the almost temperature independent $\lambda_1$ below ~ 10 K, especially when crossing $T_c$ i.e., upon entering in the superconducting phase. In conventional BCS type *s*-wave superconductors, no effects are expected in the ZF-µSR data in the superconducting electronic state. On the other hand, when TRS is broken, ZF-µSR time spectra are modified due to the appearance of spontaneous magnetic fields below $T_c$. This is typically the case for the Cooper pairs with a *p*-wave symmetry, as, *e.g.*, in $Sr_2RuO_4$ [4]. As shown in Fig. 11(b), since the time spectra do not show any relevant changes (within statistical errors), we must conclude that the TRS breaking is unlikely in the superconducting state of $Sc_5Ru_6Sn_{18}$.

Note that the TRS breaking depends on the pairing symmetry of the electrons in the superconducting state. The TRS can be broken if the superconducting state has degenerate representation, as is the case of triplet superconducting states [4]. However, in case of singlet-superconducting states, where the superconducting state has non-degenerate representations, the TRS may not be broken. While TRS breaking is possible in systems with strong spin-orbit coupling such as $Y_5Rh_6Sn_{18}$ [18]; however, it might still be conserved in systems with weak spin-orbit coupling, such as $Sc_5Ru_6Sn_{18}$.

### 3.6 Cooper pair breaking

In case of type-II superconductors, the breaking of Cooper pairs upon applying an external magnetic field may happen via two mechanisms: orbital or Pauli paramagnetic limiting field effects [34]. In case of an orbital pair-breaking mechanism, Cooper pairs break when the field-induced kinetic energy of the pair exceeds the superconducting condensation energy. On the other hand, in case of the Pauli paramagnetic effect, it is energetically favorable for electron spins to point in the direction of the applied magnetic field, thus decoupling from their partner, which results in the splitting of the Cooper pair [34]. For a BCS superconductor the orbital limit of the upper (*i.e.*, orbital) critical field is given by the Werthamer-Helfand-Hohenberg (WHH) expression [47,48]:

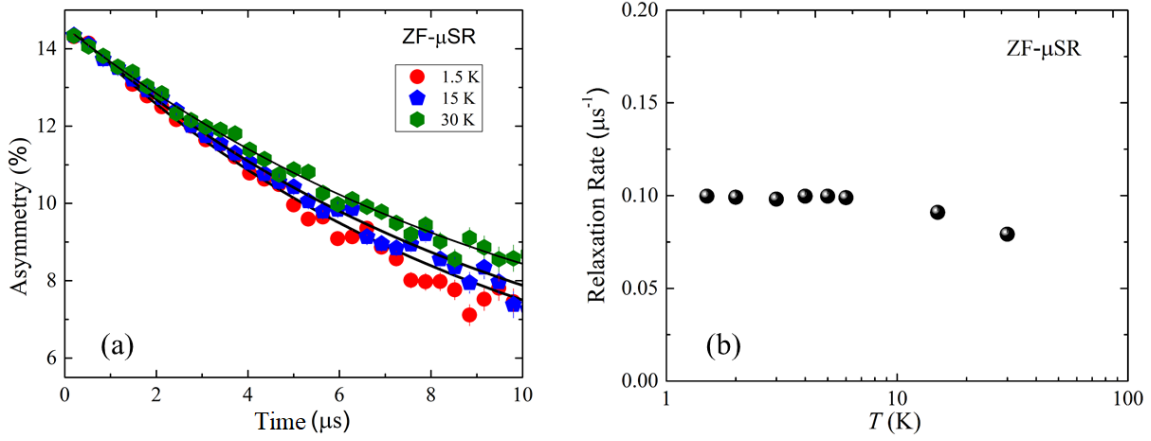

FIG. 11: (a) ZF-µSR time spectra of $Sc_5Ru_6Sn_{18}$ measured at various temperatures. The solid lines are the best-fit results by using the function, $A(T)=A_0G_{KT}(t)\exp(-\lambda_1 t)$. (b) Temperature dependence of the muon-spin relaxation rate (see text for details). The temperature independent behavior below 10 K suggests the absence of spontaneous internal magnetic fields at the muon site, indicating that time reversal symmetry is unlikely to be broken in $Sc_5Ru_6Sn_{18}$.

$$H_{c2}^{orbital}(0) = -0.693 T_c \left(\frac{dH_{c2}(T)}{dT}\right)_{T=T_c} \quad (27)$$

The slope $\left(\frac{dH_{c2}(T)}{dT}\right)_{T=T_c}$ is estimated from the $H_{c2} - T$ phase diagram (see Fig. 4) and is equal to 8.1(1) kOe / K in our case, which gives an orbital upper limiting field, $H_{c2}^{orbital}(0) \approx 19.18$ kOe. The Pauli limiting field, $H_{c2}^{P}(0)$, within the BCS theory is given by [49,50]:

$$H_{c2}^{P}(0) = 1.86\, T_c \quad (28)$$

For $T_c$ = 3.5 K, $H_{c2}^{P}(0) \approx 65.1$ kOe. The Maki parameter [51], $\alpha_M$, is used to measure the relative strength of the orbital and Pauli limiting field values and is given by the expression:

$$\alpha_M = \sqrt{2}\,\frac{H_{c2}^{orbital}(0)}{H_{c2}^{P}(0)} \quad (29)$$

From this relation, we get $\alpha_M \approx 0.29$, which means that in our case, $H_{c2}^{P}(0)$ is much larger than $H_{c2}^{orbital}(0)$ and hence implying that the upper critical field is limited by the orbital effects and the Pauli paramagnetic effect is negligible as indicated by a small value of Maki parameter. For

$Sc_5Ru_6Sn_{18}$, the calculated $H_{c2}(0)$ is close to the orbital limiting field and is much smaller than the Pauli limiting field.

## Summary

We have grown single crystals of $Sc_5Ru_6Sn_{18}$ and after succefully growing single crystals of $Sc_5Ru_6Sn_{18}$, its superconducting properties are investigated by means of powder XRD, DC and AC magnetization, specific heat and μSR measurements. The powder XRD pattern was indexed as a tetragonal structure with lattice constants, $a = 1.387(3)$ nm, $c = 2.641(5)$ nm, which imply a density of the unit cell = 7.8(3) g/cm$^3$. We find that $Sc_5Ru_6Sn_{18}$ is a type-II superconductor with $T_c \approx 3.5(1)$ K, a lower critical field $H_{c1}(0) = 157(9)$ Oe and an upper critical field, $H_{c2}(0) = 26(1)$ kOe. The zero-temperature thermodynamic critical field is estimated to be $H_c(0) = 799(6)$ Oe. With coherence length, $\xi(0) = 11.26(3)$ nm and the penetration depth, $\lambda(0) = 260(7)$ nm, $Sc_5Ru_6Sn_{18}$ has a Ginzburg Landau parameter, $k = 23(2)$. The Bean model was used to calculate the critical current density, providing $J_c(1.8$ K$) \sim 6 \times 10^8$ A/m$^2$ at 150 Oe. The de-pairing current ($J_d$) is estimated to be $3.07 \times 10^{10}$ A/m$^2$.

Our analysis of the normal-state specific heat yields a Sommerfeld coefficient, $\gamma = 36.93(6)$ mJ/mol-K$^2$, corresponding to the density of states at the Fermi level, $N(E_F) = 15.24(6)$ states/eV f.u. The superconducting transition is revealed by a sharp jump at $T_c = 3.5(1)$ K, with $\frac{\Delta C_e}{\gamma T_c} = 1.6$ for $\gamma = 36.93(6)$ mJ/mol-K$^2$, which is higher than the BCS value of 1.43 for a weakly coupled superconductor, which indicates strong electron-phonon coupling in $Sc_5Ru_6Sn_{18}$. The electronic specific heat can be fitted with a single-gap BCS model with $\Delta(0) = 0.64(1)$ meV. The Sommerfeld constant ($\gamma$) exhibits a linear variation with the applied magnetic field, indicating an s-wave superconducting pairing in $Sc_5Ru_6Sn_{18}$. TF-μSR measurements reveal that $Sc_5Ru_6Sn_{18}$ is a strongly coupled superconductor. TF-μSR measurements also suggest an s-wave character of the superconducting gap. ZF-μSR measurements do not show the presence of spontaneous internal magnetic fields and hence, indicate a preserved TRS in $Sc_5Ru_6Sn_{18}$. In table 1, we have summarized the experimentally estimated parameters of $Sc_5Ru_6Sn_{18}$, along with those of $Y_5Rh_6Sn_{18}$ [18] and $Lu_5Rh_6Sn_{18}$ [2]. It is clear that most of the parameters of $Sc_5Ru_6Sn_{18}$ are comparable with the corresponding parameters of $Y_5Rh_6Sn_{18}$ and $Lu_5Rh_6Sn_{18}$, except for $H_{c2}(0)$. Other compounds of the $R_5M_6Sn_{18}$ family are currently being investigated to clarify TRS breaking mechanism in this class of caged-type superconducting compounds.

| Parameter | Unit | $Sc_5Ru_6Sn_{18}$ | $Y_5Rh_6Sn_{18}$ | $Lu_5Rh_6Sn_{18}$ |
|---|---|---|---|---|
| $T_c$ | K | 3.5(1) | 3 | 4(1) |
| $H_{c2}(0)$ | kOe | 26(1) | 43 | 72.4 |
| $\gamma$ | mJ/mol-K$^2$ | 36.93(6) | 38.13(3) | 48.1(5) |
| $\Delta C_e/\gamma T_c$ | | 1.6 | 1.95 | 2.06 |
| $2\Delta(0)/k_B T_c$ | | 4.25(4) | 4.26(4) | 4.26(4) |
| $\theta_D$ | K | 205(1) | 183(2) | 157(2) |
| $m^*$ | | 1.64(4)$m_e$ | 1.21$m_e$ | 1.32$m_e$ |
| $n_s$ | carriers/m$^3$ | 7.05(2) × 10$^{26}$ | 2.3 × 10$^{28}$ | 2.6 × 10$^{26}$ |
| $H_{c1}(0)$ | Oe | 157(9) | | |
| $H_c(0)$ | Oe | 799(6) | | |
| $J_c(1.8\ K)$ | A/m$^2$ | 6(3)×10$^8$ | | |
| $J_d$ | A/m$^2$ | 3.07(5)×10$^{10}$ | | |
| $\lambda(0)$ | nm | 260(7) | | |
| $\xi(0)$ | nm | 11.26(3) | | |
| $k_{GL}$ | | 23(2) | | |
| $\lambda_{eff}(0)$ | nm | 774(8) | | |
| $N(E_F)$ | states/eV $f.u.$ | 15.24(6) | | |
| $l$ | nm | 8.14(5) | | |

Table 1: Superconducting parameters of $Sc_5Ru_6Sn_{18}$, $Y_5Rh_6Sn_{18}$ [18] and $Lu_5Rh_6Sn_{18}$ [2].

**Acknowledgments**

We acknowledge the Ministry of Science and Technology (MOST) of Taiwan, which supported the work, via the grants: MOST 104-2221-M-006-010-MY3, MOST 104-2119-M-006-017-MY3. DK would like to acknowledge MOST for the postdoctoral fellowship under project grant number MOST 105-2811-M-006-022. TS and TS acknowledge support from SNF – the Swiss National Science Foundation. Part of this work is based on experiments performed at the Swiss Muon Source SµS, Paul Scherrer Institute, Villigen, Switzerland. We thank M R Lees for fruitful discussion.